# Room-temperature strong coupling in a single photon emitter-dielectric metasurface system

T. Thu Ha Do,[1][†] Milad Nonahal,[2,3][†] Chi Li,[2] Vytautas Valuckas,[1] Arseniy I. Kuznetsov,[1] Hai Son Nguyen,[4,5*] Igor Aharonovich,[2,3*] Son Tung Ha[1*]

[1]Institute of Materials Research and Engineering, A*STAR (Agency for Science, Technology and Research), 138635 Singapore.

[2]School of Mathematical and Physical Sciences, Faculty of Science, University of Technology Sydney, Ultimo, New South Wales 2007, Australia.

[3]ARC Centre of Excellence for Transformative Meta-Optical Systems, University of Technology Sydney, Ultimo, New South Wales 2007, Australia.

[4]Univ Lyon, Ecole Centrale de Lyon, CNRS, INSA Lyon, Universite Claude Bernard Lyon 1, CPE Lyon, CNRS, INL, UMR5270, 69130 Ecully, France.

[5]Institut Universitaire de France (IUF), F-75231 Paris, France

*Corresponding authors. Email: hai-son.nguyen@ec-lyon.fr, igor.aharonovich@uts.edu.au, ha_son_tung@imre.a-star.edu.sg

[†]These authors contributed equally to this work.

**Abstract:** Single-photon sources with high brightness and long coherence time are promising qubit candidates for quantum technology. To this end, interfacing emitters with high-finesse cavities is required, especially in the strong coupling regime, which so far has only been limited to cryogenic temperatures. Here, we experimentally demonstrate, at room temperature, strong coupling between a single photon emitter and a novel cavity based on optical bound states in the continuum. A remarkably large Rabi splitting of ~4 meV is achieved thanks to the combination of the narrow linewidth and large oscillator strength of emitters in hexagonal boron nitride and the efficient photon trapping of the cavity. Our findings unveil new opportunities to realise scalable quantum devices and explore fundamentally new regimes of strong coupling in quantum systems at room-temperature.

**One-Sentence Summary:** Room-temperature strong coupling of a single-photon emitter and a metasurface-based cavity is reported.

Light-matter interaction at single-photon levels is technologically important for quantum communication, information and quantum computing. Single-photon emitters (SPEs) can act as elementary blocks (qubits) for solid-state quantum computers, and their interaction can be mediated by cavity modes (1). In most cases, SPE-cavity systems undergo incoherent processes of weak coupling, in which the fluorescence is enhanced by the Purcell effect. On the other hand, in the strong coupling regime, emitters and photons exchange energy coherently, leading to two new hybrid half-light-half-matter states (polaritons) separated by a characteristic Rabi splitting energy $2g$, where $g$ is the coupling strength. The relationship between $g$ and the optical transition dipole moment $\boldsymbol{\mu}$, local electric field $\boldsymbol{E}$, oscillator strength $f$ and cavity mode volume $V$ can be expressed as:

$$g = \boldsymbol{\mu}.\boldsymbol{E} = \hbar \sqrt{\frac{\pi e^2 f}{4\epsilon_r \epsilon_0 m_0 V}} \qquad (1)$$

where $\epsilon_r \epsilon_0$ is the dielectric constant of the cavity material and $m_0$ is the free electron mass. Theoretically, the condition $g \geq |\kappa_{cav} - \kappa_{SPE}|/2$, where $\kappa_{cav}$ and $\kappa_{SPE}$ are dissipative decay rates of the cavity and the emitter, respectively, guarantees real solutions for two hybrid eigenstates (2). This requirement can be fulfilled by (i) enhancing $g$ via reducing $V$, increasing $f$ and/or (ii) lowering the critical threshold for coupling strength via using narrow-line emitters (small $\kappa_{SPE}$) and high-finesses cavities (high quality ($Q$) factor, small $\kappa_{cav}$). Since transition dipole moment of single emitters is generally small, the coupling strength $g$ of SPE-cavity systems is also small.

Substantial efforts have been devoted to fabricating optical cavities sustaining both high-$Q$ and small-$V$ simultaneously (3-7). However, it consequently imposes a serious challenge for precisely positioning a single emitter at the electric field maxima within such small active mode volumes and aligning emitter dipole moments with cavity fields (8). Strong coupling emission of SPEs has been experimentally observed in high-$Q$ cavities at cryogenic temperature (9-13). Nevertheless, it remains challenging at elevated temperatures when the radiation loss $\kappa_{SPE}$ is substantially enhanced due to the interaction with the phonon bath. Therefore, the design of optical cavities and the choice of SPE sources become critical to satisfy the strong coupling criteria.

Here, we demonstrate room-temperature strong coupling of SPEs in a high-$Q$, large-$V$ cavity using the concept of bound state in the continuum (BIC). In theory, photons are perfectly trapped inside BIC cavities ($\tau_{BIC} \rightarrow \infty$) and the condition reduces to $g \geq \kappa_{SPE}/2$, since $\kappa_{BIC} = \tau_{BIC}^{-1}$ equals 0. Therefore, the strong coupling is no longer restricted by the cavity loss. Such a high $Q$-factor of BIC plays a vital role in the recent progress in nonlinear optics (14-16), functional metasurface (17), lasing (18-21), and Bose-Einstein condensation (22). SPEs used in this work are carbon-based colour centres generated in a few-layer-thick hexagonal boron nitride (hBN) film, which are reported to have bright emission, non-blinking nature, ultrahigh Debye-Wallor factor ($F_{DW}$ ~0.82), optically addressable spin states, and high single-photon purity at temperatures up to 800 K (23-25). Most importantly, their emission has uniquely narrow linewidth at room temperature, which can even reach the Fourier-transform limit (i.e., $\kappa_{SPE}$ ~0.2 μeV) (26), greatly favouring for strong coupling. In our SPE-BIC systems, we resolved a Rabi splitting of ~4 meV at room temperature, which is unprecedented at single-photon emitter levels. Our results strongly suggest that the combination of BIC cavities and SPEs in two-dimensional materials is a viable route towards scalable quantum-based devices operating at ambient conditions.

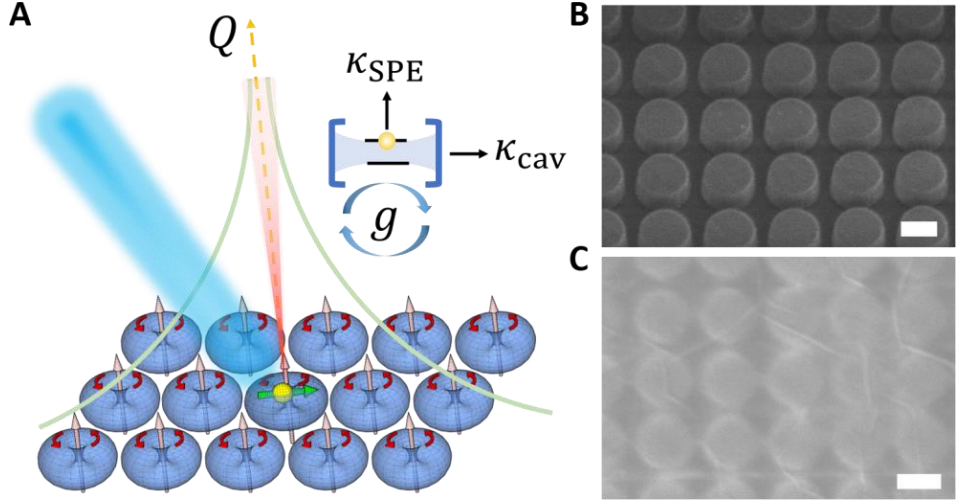

**Fig. 1. Single-photon emitter in the bound-state-in-the-continuum cavity.** (**A**) Illustration of coupling between a SPE and a BIC cavity. The BIC mode is formed by an array of vertical magnetic dipoles (pink arrows), which associates with an electric field (blue doughnuts) circulating in the *x-y* plane (red arrows). An SPE with a horizontal transition dipole moment (yellow sphere with green arrow) spatially overlaps with the electric-field hotspots. (**B**) 30° tilted SEM image of the fabricated TiO$_2$ nanopillar array ($D \sim 260$ nm, gap $\sim 40$ nm). (**C**) SEM image of the array after transferring a 3-nm thick hBN film on top. Scale bars in (**B**) and (**C**) are 200 nm.

Figure 1A illustrates the concept of coupling between a single-photon emitter and a BIC cavity formed by a sub-diffractive array of vertical magnetic dipoles. The SPE (yellow sphere) is located within the electric field (***E***) distribution (blue doughnuts), and the transition dipole moment (***μ***) of the emitter (green arrow) is aligned with ***E***. Hence, the coupling strength $g \propto \boldsymbol{\mu}.\boldsymbol{E}$ is optimal. In particular, our BIC cavity is constructed by a 50×50 μm$^2$ square array of nanopillars made of titanium dioxide (TiO$_2$) as shown in the scanning electron microscope (SEM) image in Fig. 1B. Being a relatively high refractive index (i.e., $n \sim 2.5$ at 600 nm) and lossless material at optical frequencies, TiO$_2$ nanostructures manifest themself as ideal candidates for an ultrahigh-$Q$ cavity that have been extensively demonstrated to support BIC modes originated from vertical electric dipoles, vertical magnetic dipoles and electric quadrupoles (*20*). Here, the nanopillar diameter ($D$) and their gap are varied to tune the resonance position of the BIC mode to spectrally overlap with the emission from SPEs. A 3-nm-thick hBN film (Fig. S1) was grown by metal-organic vapour phase epitaxy on a sapphire substrate and then transferred onto the fabricated TiO$_2$ nanostructures (Fig. 1C). The SPEs were carbon-related defects in hBN generated by thermal annealing (see Supplementary Materials, Methods).

Figure 2A shows the measured (left) and simulated (right) angle-resolved reflection spectra of the TiO$_2$ nanopillar array with $D \sim 260$ nm and gap $\sim 40$ nm. The spectral narrowing and the vanishing of reflectance signal when approaching normal incidence ($\theta = 0°$) indicate the formation of the symmetry-protected BIC at $E_{BIC} \sim 2.107$ eV. The multipolar analysis reveals that this BIC mode originates from the vertical magnetic dipole, as reported elsewhere (*20*).

The electric field associated with the vertical magnetic dipole is concentrated within the nanopillars' peripheries in the *x-y* plane with the maximal field intensity at ~75 nm from the centre (Fig. 2B). Such circulating **E**-field enables maximum coupling with the in-plane dipoles. Furthermore, the field extends ~40 nm outside of the nanopillar surface along the $z$-direction (Fig. 2C), ensuring spatial overlap with SPEs in the 3-nm thick hBN film. In contrast to other high-$Q$ dielectric cavities, at BIC conditions, electric-field hotspots exist on every nanopillar, increasing the possibility of spatial overlap with randomly distributed SPEs. The experimental angle-dependent energy $E(\theta)$ and full-width at half-maximum (FWHM) $\kappa_{BIC}(\theta)$ of the BIC band extracted from measured reflection are shown in Fig. 2D. The fitting values (solid lines) are calculated assuming that the cavity is lossless (i.e., $\kappa_{BIC}$ (0°) = 0 meV).

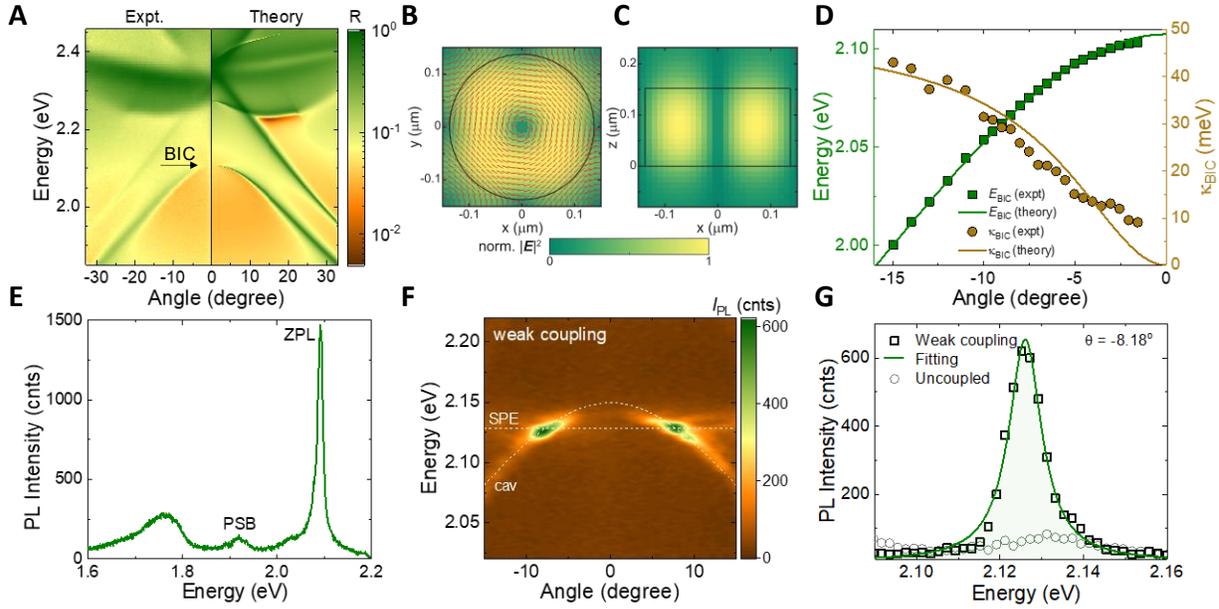

**Fig. 2. Optical characterisation for the BIC cavity, SPE and their weak coupling.** (**A**) Experimental (left) and calculated (right) angle-resolved unpolarised reflectance spectra of the designed TiO$_2$ nanopillar array showing vertical magnetic-dipole BIC resonance at $E_{BIC}$~ 2.107 eV at $\theta = 0°$. (**B**) Top view and (**C**) side view of the electric field distribution at the BIC frequency in the nanopillars. (**D**) Experimental resonance energies (green squares) and FWHM (light-brown circles) of the BIC mode observed in (**A**) as a function of $\theta$, and the corresponding fitting curves (coloured solid lines). (**E**) Typical PL spectrum of carbon-related SPEs showing the zero phonon line (ZPL) peak at ~2.092 eV, and a phonon side band (PSB) at ~1.919 eV. (**F**) Angle-resolved photoluminescence (PL) spectra in a weak coupling regime, showing a PL enhancement of the SPE. The cavity mode and the SPE PL position are shown by the white dashed lines. (**G**) PL spectra of the coupled SPE extracted from (**F**) at -8.18º (squares) and the uncoupled SPE (i.e., circle) under the same measurement conditions.

Figure 2E shows a typical SPE photoluminescence (PL) spectrum at room temperature with a zero-phonon line (ZPL) at $E_{SPE}$ ~2.092 eV. Upon scanning over an area of 50×50 µm² on the as-grown hBN film, we identified 36 individual single photon emitters (i.e., second-order correlation function at zero-time delay $g^2(0)$) < 0.5) with ZPL position ranging from ~2.026 eV to ~2.149 eV (Fig. S2). The low SPE density (~$10^7$ cm$^{-2}$) increases the chance of isolating a single SPE from the surrounding SPEs with similar energies. Furthermore, dipole moments of the SPEs have been previously shown to be oriented in the plane of the hBN film (27), ensuring the coupling with the circulating electric field of the vertical magnetic dipole BIC mode ($\boldsymbol{\mu.E} > 0$).

We study the coupling of SPE and BIC modes by performing angle-resolved and energy-resolved PL measurements under non-resonance excitation ($E_{\mathrm{Exc}}$ ~2.541 eV). Figure 2F shows the angle-resolved PL spectra of a SPE emitting at ~2.129 eV, weakly coupled with the BIC mode ($E_{\mathrm{BIC}}$ ~2.149 eV) in a nanopillar array with $D$ ~275 nm and gap ~50 nm. Noticeably, no spectral splitting has been observed, but rather a significant enhancement of PL intensity from the coupled SPE compared to the uncoupled one (Fig. 2G). Furthermore, the BIC band dispersion remains unchanged compared to the uncoupled cavity mode (Fig. S3A). When there is a large mismatch between $E_{\mathrm{SPE}}$ and $E_{\mathrm{BIC}}$ (Fig. S4), only weak coupling cases (i.e., no Rabi splitting) can be observed.

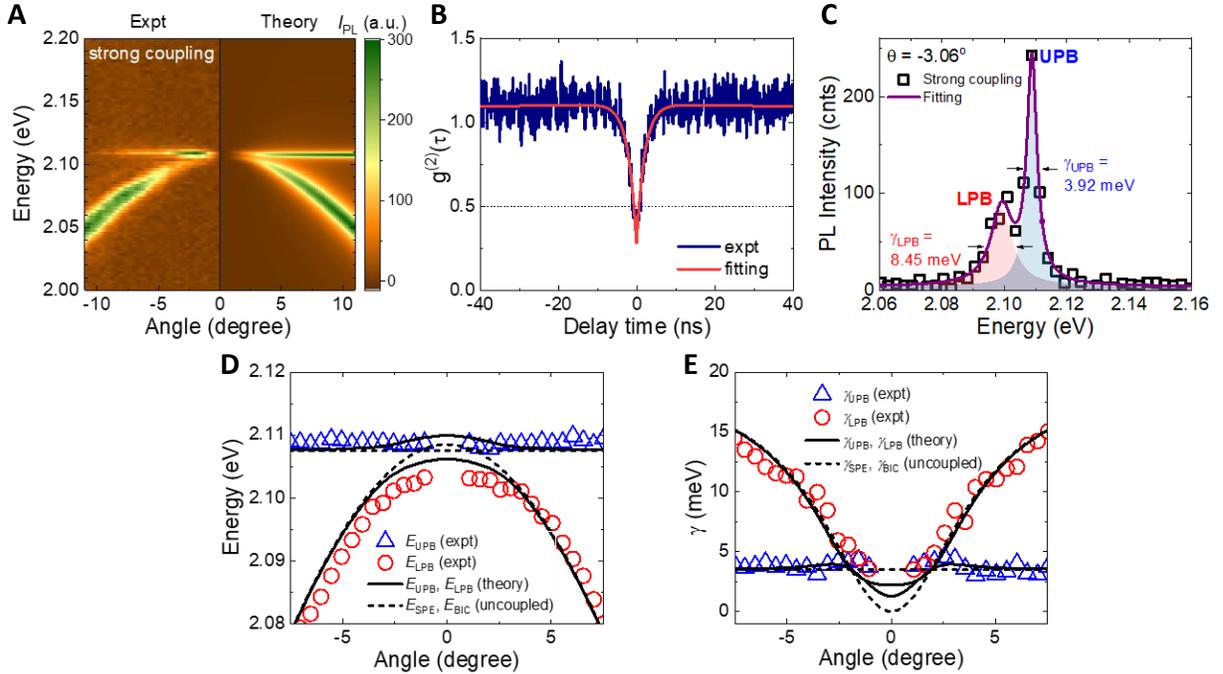

**Fig. 3. Room-temperature strong coupling in SPE-BIC systems.** (**A**) Experimental angle-resolved PL spectra (with excitation power $P_{\mathrm{exc}}$ ~ 1.329 kW/cm$^2$) (left) and the simulation (right) in a strong coupling regime, evidenced by the clear spectral splitting and vanishing emission at normal incidence on both polariton branches. (**B**) Antibunching dip $g^2(0) = 0.28$ confirming the single-photon emission nature at the measured strong coupling spot. (**C**) PL spectra extracted at $\theta = -3.06°$ (black squares) and the double-Lorentzian fitting (purple curve). (**D**) Energy and (**E**) FWHM of the upper (blue up-triangles) and lower (red circles) polariton branches (i.e., UPB and LPB, respectively) at different angles and the theoretical fitting using coupled oscillator model (solid black lines). The energies and FWHM of the uncoupled SPE and BIC modes are shown by dashed black lines.

In contrast, when the $E_{\mathrm{SPE}}$ and $E_{\mathrm{BIC}}$ match (i.e., TiO$_2$ array with $D$ ~260 nm and gap ~40 nm), a clear spectral splitting of the upper and lower branches can be seen, as shown in Fig. 3A. Importantly, both branches show no coupling to the radiation field at the normal incidence that inherits from the symmetry-protected BIC nature. The single-photon emission purity of the investigated emitter at the measured strong coupling spot is confirmed by the antibunching dip $g^2(0) = 0.28$ (Fig. 3B). The energy splitting between two BIC polaritons is clearly seen in the PL spectrum near normal incidence ($\theta \sim$ -3.06°), which is fitted by a double-Lorentzian function (Fig. 3C). Furthermore, the polariton dispersions are modified from the original uncoupled cavity dispersion that is revealed by comparing the cross-sectioned PL spectra in the two cases (Fig. S3B).

The extracted polariton energies and linewidths are plotted against the incident angle, as shown in Fig. 3D and Fig. 3E, respectively. The FWHM ($\gamma$) of the two polariton bands reduces and becomes identical when $|\theta| \sim$1.93°. The BIC polariton properties can be well modelled by using coupled oscillator theory (see Supplementary Text). The calculated eigenvalues are shown by black lines, where the real parts are eigenenergies (Fig. 3D), and imaginary parts are FWHM of polariton branches (Fig. 3E). We used a detuning energy of 1 meV between the BIC resonance and the SPE at $\theta = 0$. The Rabi splitting is then defined to be $\sim$4 meV at $|\theta| \sim$1.24°, corresponding to a coupling strength $g \sim$2 meV. It is noted that the vanishing of the signal at normal emission in both branches is the hallmark of polariton BIC that has been previously reported for the strong coupling regime between excitons in quantum wells and photonic BIC (*22,28*). The spectral splitting, the narrowing of linewidth, the vanishing of the signal at normal incidence and the good agreement between experiment and theory confirm the strong coupling nature of the SPE and BIC in our systems.

Remarkably, the coupling strength $g \sim$2 meV in our system is one order of magnitude higher than the record established for the strong coupling regime using InAs/GaAs quantum dots at a cryogenic temperature (*9-13*). Using equation (1), we deduce an oscillator strength per unit volume $f/V$ of two orders of magnitude higher than the one in InAs/GaAs systems. One possible way to acquire such a large $f$ is through the interaction of BIC modes with other midgap states via the polariton branches as theoretically proposed (*29*). Nevertheless, further study is needed to unravel the underlying physics behind the large oscillator strength in this class of SPEs.

To further investigate the SPE-BIC coupling, we employed excitation power to detune the energy difference between the SPE and BIC mode. Specifically, when increasing the power density from 0.672 kW/cm$^2$ to 25.784 kW/cm$^2$, the BIC resonance energy increases by $\sim$2.2 meV, while the SPE peak remains unchanged (Fig. S5). Therefore, the excitation power can be used to vary the detuning energy between SPE and BIC. Here, the detuning at $\theta = 0$ is varied from 1.0 meV to 3.8 meV when increasing the excitation power. It is worth noting that the SPEs are optically stable under our experimental conditions.

As a result of the varying detuning energy, we observe the energy shift of polariton branches, as shown in Fig. 4A for two representative excitation powers $P_{exc} \sim$1.329 kW/cm$^2$ and $P_{exc} \sim$25.784 kW/cm$^2$. The power-dependent PL spectra extracted at an angle close to $\theta = 0$ collected at various excitation powers are shown in Fig. 4B, and the corresponding energy shifts of upper polariton branch (UPB) and lower polariton branch (LPB) are plotted as a function of excitation power in Fig. 4C. With increasing the excitation power, both polariton branches show blueshifts. Over the investigated power range, the two branches do not cross each other. At the highest power used, the UPB and LPB peak shifts by $\Delta E_{UPB} \sim$ 3.0 meV and $\Delta E_{LPB} \sim$ 0.8 meV, respectively. The observed energy shifts can be well reproduced by the coupled oscillator model (solid curves) with the power-induced changes in the detuning energy. The power-dependence of $E_{UPB}$ and $E_{LPB}$ further supports our attribution of the band splitting to the strong coupling. Indeed, the hybrid nature of mixed BIC-SPE in the two bands imposes the blueshift for both bands, with the shifting dictated by the BIC fraction.

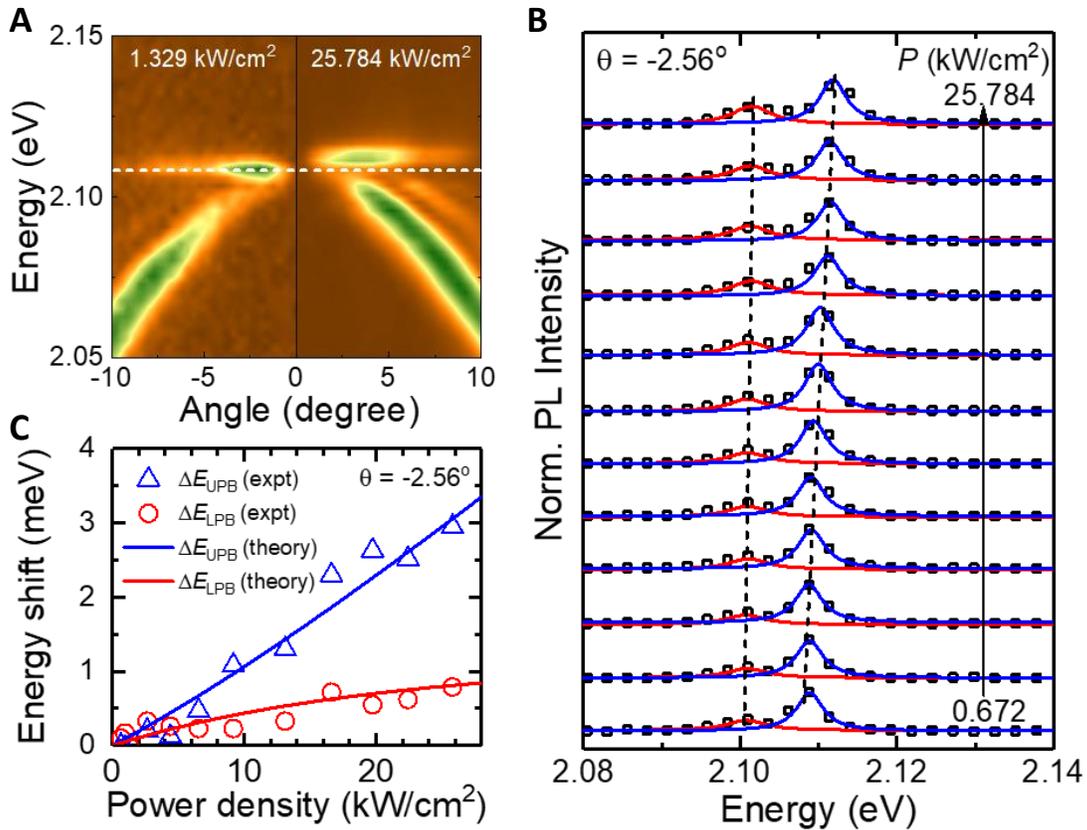

**Fig. 4. Tuning of strong coupling by excitation power.** (**A**) Angle-resolved PL spectra acquired under different excitation power: $P_{exc}$~1.329 kW/cm² (left) and $P_{exc}$ ~25.784 kW/cm² (right) showing the blueshifts of both polariton branches with increasing power. The white dashed line is guideline. (**B**) Normalised PL spectra extracted at $\theta = -2.56°$ under varying excitation power density, which induces changes in detuning energy and energy splitting. (**C**) Energy shift of UPB (blue triangles) and LPB (red circles) relative to the lowest used power. Theoretical fitting curves are shown by solid lines considering the power-induced energy shift of the uncoupled SPE and BIC modes.

In conclusion, we have experimentally realised, at room temperature, strong coupling between a single-photon emitter and a bound-state-in-the-continuum cavity. The absence of emission at normal incidence, along with the narrowing of bandwidths at BIC condition and the characteristic Rabi splitting between the BIC polariton branches of up to ~4 meV, were resolved. We exploited the versatility of the BIC mode to overcome the current challenging problems in controlling the position and dipole orientation of the quantum emitters, while the ultrahigh quality factor of the BIC cavity and the narrow linewidth of SPEs make strong coupling possible. We expect that our findings will open up a rich vein of research on the manipulation of emission properties from single-photon emitters. Future work on the deterministic positioning of single-photon sources at designated locations will allow for the study of quantum entanglement and quantum coherence that have important implications in nonlinear quantum optics, integrated nanophotonics and quantum information processing.

**Acknowledgments:** T. T. H. D., V. V., A. I. K., and S. T. H. gratefully acknowledge the financial support from AME Yong Individual Research Grant (A2084c0177) and A*STAR MTC-Programmatic Fund (M21J9b0085). I. A. acknowledges the financial support from the Australian Research Council (CE200100010) and the Office of Naval Research Global (N62909-22-1-2028). H. S. N acknowledges French National Research Agency (ANR) under the project POPEYE (ANR-17-CE24-0020), the IDEXLYON from Université de Lyon, Scientific Breakthrough project TORE within the Programme Investissements d'Avenir (ANR-19-IDEX-0005), and the Auvergne- Rhône-Alpes region in the framework of PAI2020.

**Funding:**

AME Yong Individual Research Grant (A2084c0177)

A*STAR MTC-Programmatic Fund (M21J9b0085)

Australian Research Council (CE200100010)

Office of Naval Research Global (N62909-22-1-2028)

Project POPEYE (ANR-17-CE24-0020)

Programme Investissements d'Avenir (ANR-19-IDEX-0005)

**Author contributions:** S. T. H. and I. A. conceived the project. T. T. H. D. and S. T. H. fabricated the dielectric nanoantenna. V. V. characterised the nanostructures. M. N. and C. L. prepared the few-layer hBN samples with single-photon emitters. T. T. H. D., M. N., C. L., and S. T. H. performed optical measurements and analysed the data. H. S. N. performed theoretical calculations. A. I. K. discussed the data and manuscript. T. T. H. D. wrote the manuscript with input from all co-authors. T. T. H. D. and M. N. contributed equally to this work.

**Competing interests:** Authors declare that they have no competing interests.

**Data and materials availability:**

All data are available in the main text or the supplementary materials.

**Supplementary Materials**

Methods

Supplementary Text

Figs. S1 to S5

References (*30–32*)

# Supplementary Materials for

## Room-temperature strong coupling in a single photon emitter-dielectric metasurface system

T. Thu Ha Do,[1][†] Milad Nonahal,[2,3][†] Chi Li,[2] Vytautas Valuckas,[1] Arseniy I. Kuznetsov,[1] Hai Son Nguyen,[4,5][*] Igor Aharonovich,[2,3][*] Son Tung Ha[1][*]

*Correspondence to: hai-son.nguyen@ec-lyon.fr, igor.aharonovich@uts.edu.au, ha_son_tung@imre.a-star.edu.sg

**This PDF file includes:**

> Methods
> Supplementary Text
> Figs. S1 to S5
> References and Notes (*30-32*)

**Methods**

<u>Nanofabrication</u>

A 150-nm thick film of TiO$_2$ was deposited onto a quartz substrate by ion-assisted deposition (IAD, Oxford Optofab3000), and then a 30-nm-thick film of Cr was deposited by electron-beam evaporation (Angstrom EvoVac) as a hard mask layer. A negative electron-beam resist, hydrogen silsesquioxane (HSQ, Dow Corning), was spin-coated on TiO$_2$/Cr/quartz at 5000 rpm for 60 seconds, followed by baking at 120°C for 2 minutes and at 180°C for 2 minutes. The patterning was done with electron beam lithography (Elionix ELS 7000), and the sample was developed using a salty developer (i.e., solution of 1 wt. % NaOH and 4 wt.% NaCl in DI water) for 4 minutes and rinsed with DI water (*30*). The HSQ pattern was first transferred to Cr layer by using inductively coupled plasma reactive ion etching (ICP-RIE, Oxford Plasmalab 100) with a mixture of Cl$_2$ gas at 15 standard cubic centimetres per minute (sccm) and O$_2$ (2 sccm) gas at 10 mTorr and 7°C. After that, Cr pattern was transferred to TiO$_2$ film by using CHF$_3$ gas using the same etching machine at 25 sccm, 32 mTorr and 5°C. Finally, the whole sample was immersed in a chromium etchant solution (Merck) for 15 minutes to remove the remaining Cr and resist layers. The final sample was then rinsed in DI water and blow-dried with nitrogen gas.

<u>hBN film growth process</u>

A few-layer hBN film was grown on a sapphire substrate by metal-organic vapour-phase epitaxy (MOVPE) following the method reported previously (*31*). More specifically, triethyl boron (TEB) and ammonia served as boron and nitrogen precursors, respectively. The sapphire substrate was initially functionalised with ammonia prior to hBN growth at a temperature of about 1,000°C. The chamber temperature was subsequently elevated to 1,350°C, and the precursors were introduced into the chamber once the temperature was stabilised. The precursors were intentionally injected into the reactor with short pulses lasting 1 to 3 seconds to minimise parasitic reactions.

<u>hBN film transfer process</u>

Milimeter-sized hBN film was transferred from the growth sapphire substrate onto BIC structures using a wet transfer method. Approximately a 300-nm PMMA (A4, Mircochem) layer was spin-coated onto hBN/sapphire substrate and baked at 120°C for 3 mins to evaporate the polymer solvent. The sample was then floated onto a 1M KOH aqueous solution to etch the sapphire interfacial layer and detach it from the hBN/PMMA film. After picking the floating film with the targeted substrate, the film was washed three times with DI water to remove the remaining base. A similar process was done to transfer hBN film onto the BIC structures. To gently remove water residue without the formation of wrinkles and bubbles, the substrate was placed in a vacuum desiccator for 30 minutes, followed by heating on a hotplate for another 30 minutes. This step is critical since some wrinkles are thick enough to interfere with the BIC modes. Thereafter, PMMA film was removed in a warm acetone bath overnight and then the sample was cleaned by a UV ozone (ProCleaner™ Plus, Bioforce Nanosciences Inc.) for 10 mins to remove the remaining polymer. Finally, the hBN/BIC sample was annealed on a hotplate at 500°C for 2 hours before measurements to enhance adhesion between the film and substrate.

<u>Photoluminescence and photon autocorrelation measurements</u>

The optical measurements of hBN SPEs were carried out on a home-built confocal microscope. We used a 532 nm continuous-wave laser for the excitation. Laser scanning was manipulated by an X-Y scanning mirror (FSM300™, Newport Corp.). A 100x objective (Nikon, 0.9 NA) and one green dichroic mirror were used for the collection. Reflected laser and PL signals were filtered with an extra 568-nm long-pass filter. An additional bandpass filter centred at 587 nm with 15 nm bandwidth was used to only collect the emission from SPEs near resonance with the BIC mode. For detection, a flip mirror was used to guide the signal into a spectrometer (Princeton Instruments Inc.) or a single-mode fibre connected with two avalanche photodiodes (APDs) (Excelitas Technologies) via a 50:50 fibre splitter. For the photon antibunching characterisation, the timing and correlation between two APDs were done by a correlator module (PicoHarp300[TM]). We recorded the coincidence count histogram with a 64-ps bin width resolution.

<u>Angle-resolved optical characterisations</u>

Angle-resolved spectroscopic measurements were performed using an inverted optical microscope (Nikon Ti-U) coupled to a spectrometer equipped with an electron-multiplying charged-coupled detector (EMCCD, Andor Newton 971). For reflectance measurements, light from a halogen lamp was focused onto the sample surface via a 50× objective (Nikon, NA = 0.55) with a spot size of ~7 µm. For photoluminescence, the excitation was a 488 nm continuous-wave laser. The signals were collected by the same objective and passed through a series of lenses for back focal plane imaging. The wavelengths were resolved by a spectrograph (Andor SR-303i) with a single grating groove density of 150 gr/mm and a slit size of 100 µm.

**Supplementary Text**

<u>RCWA simulation</u>

The RCWA simulation for the field distribution was performed using the S$^4$ package provided by the Fan Group at the Stanford Electrical Engineering Department (*32*). The complex refractive index of TiO$_2$ is obtained from ellipsometric measurements.

<u>BIC models</u>

The energy and losses of the BIC mode as functions of angle are modelled by:

$$E_{\text{BIC}}(\theta) = E_{\text{BIC}}(\theta = 0) + U - \sqrt{U^2 + v^2 . k(\theta)^2}$$

and

$$\kappa_{\text{BIC}}(\theta) = \left(\frac{1}{\kappa_\infty} + \frac{1}{\alpha . k(\theta)^2}\right)^{-1}$$

where $k(\theta) = (2\pi/\lambda) \sin \theta$ is the in-plane wavevector, $v$ represents the group velocity of the BIC mode at high oblique angles, $\kappa_\infty$ is the losses of the BIC mode at high oblique angles. Importantly, the dependence $\alpha . k(\theta)^2$ represents the transformation from BIC at $\theta = 0$ to quasi-BIC when a slightly oblique angle is introduced. Here, the experimental results are fitted using $U = 0.3$ eV, $v = 0.1$ eV.µm, $\kappa_\infty = 20$ meV and $\alpha = 30$ eV.µm$^2$.

<u>Coupled oscillator models</u>

The light-matter interaction is expressed as:

$$H = \begin{pmatrix} E_{\text{BIC}}(\theta) + i\kappa_{\text{BIC}}(\theta)/2 & g \\ g & E_{\text{SPE}}(\theta) + i\kappa_{\text{SPE}}/2 \end{pmatrix},$$

where $E_{\text{BIC}}(\theta)$ and $E_{\text{SPE}}(\theta)$ are energy dispersion of the uncoupled BIC cavity and the SPE; $\kappa_{\text{BIC}}(\theta)$ and $\kappa_{\text{SPE}}$ represent loss of BIC and SPE modes (FWHM in emission spectra); $g$ is the coupling strength.

The complex eigenenergies of the Hamiltonian are:

$$\tilde{E}_{\pm}(\theta) = \frac{E_{\text{BIC}}(\theta) + E_{\text{SPE}}(\theta)}{2} + i\frac{\kappa_{\text{BIC}}(\theta) + \kappa_{\text{SPE}}}{4} \pm \sqrt{\Delta(\theta)^2 + g^2}$$

with $\Delta(\theta) = \frac{E_{\text{BIC}}(\theta) - E_{\text{SPE}}(\theta)}{2} + i\frac{\kappa_{\text{BIC}}(\theta) - \kappa_{\text{SPE}}}{4}$ .

The energies and linewidths of the polariton states are given by the real and imaginary components of the eigenenergies:

$$E_{\text{UPB/LPB}}(\theta) = \text{Re}[\tilde{E}_{\pm}(\theta)]$$

$$\gamma_{\text{UPB/LPB}} = 2\text{Im}[\tilde{E}_{\pm}(\theta)]$$

The polariton states are eigenvectors of the Hamiltonian, given by:

$$|\text{UPB/LPB}\rangle = |\text{BIC}\rangle + A_{\pm}(\theta)|\text{SPE}\rangle$$

with $A_{\pm}(\theta) = \pm\sqrt{1 + \left[\frac{\Delta(\theta)}{g}\right]^2} - \frac{\Delta(\theta)}{g}$

The photonic (BIC) and excitonic (SPE) fractions of the polariton states are given by:

$$W_{\text{UPB/LPB}}^{BIC}(\theta) = \left(1 + \left|A_{\pm}(\theta)\right|^2\right)^{-1}$$

$$W_{\text{UPB/LPB}}^{SPE}(\theta) = 1 - W_{\text{UPB/LPB}}^{BIC}(\theta)$$

A blueshift $\delta_{BIC}$ of the BIC mode will lead to blueshifts of the polariton states that are given by:

$$\delta_{\text{UPB/LPB}}(\theta) = \frac{\delta_{BIC}}{2}\left\{1 \mp \left(\left[\frac{g}{\Delta(\theta)}\right]^2 + 1\right)^{-1/2}\right\}$$

One may show that the blueshift increases with superior BIC fraction in the polaritonic state.

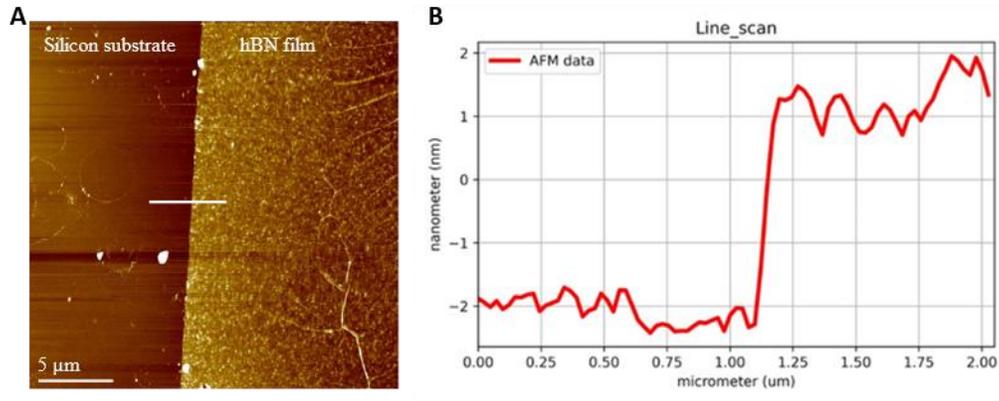

**Fig. S1.** (**A**) AFM image of the as-grown hBN film on a sapphire substrate before transferring onto TiO₂ nanostructures. (**B**) Height profile extracted along the white line in (**A**) revealing the film thickness of ~3 nm.

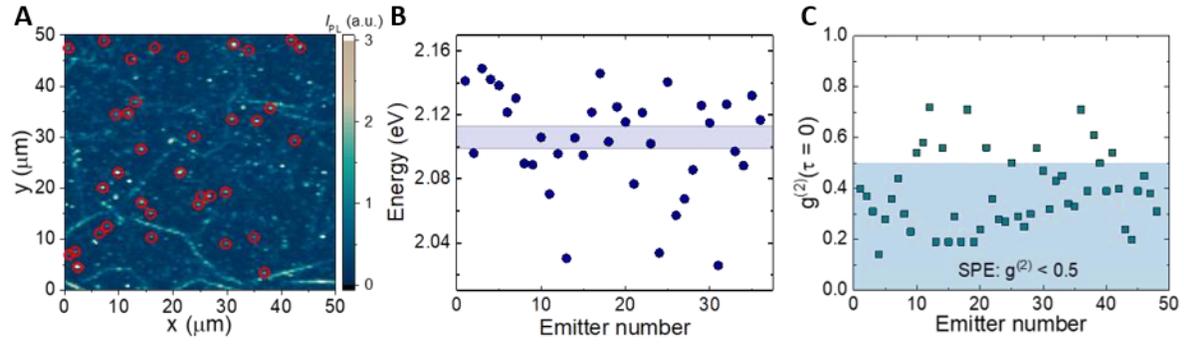

**Fig. S2.** (**A**) Confocal PL map taken in a region of 50×50 μm² on an hBN film, where 36 SPEs (with $g^2(0) < 0.5$) were identified (red circles). (**B**) Statistics of emission energies showing only 4 (out of 36) SPEs emitting at ~2.106 eV (blue shaded-area), which match with the BIC resonance presented in Fig. 3 for strong coupling. (**C**) Statistics of $g^2(0)$ for all emitters detected by PL confocal setup. Single-photon emitters are defined as $g^2(0) < 0.5$ (shaded area).

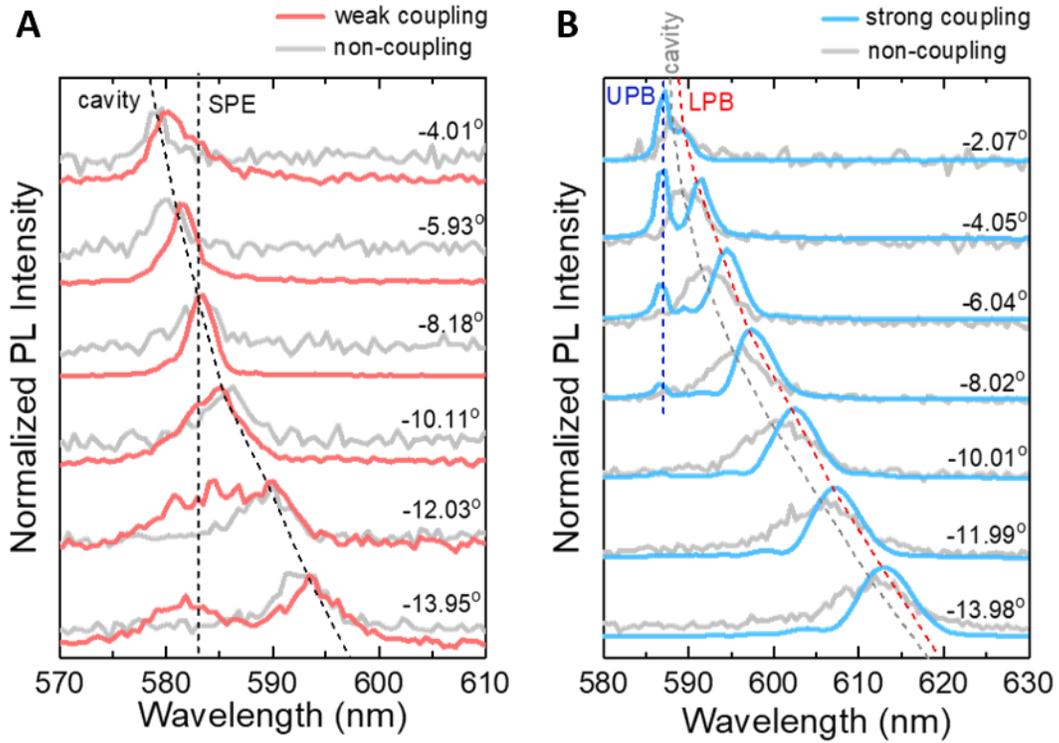

**Fig. S3.** Comparison of PL spectra extracted at different angles for (**A**) weak coupling (solid red lines) versus non-coupling (solid grey lines) cases and (**B**) strong coupling (sky-blue solid lines) versus non-coupling (solid grey lines). In the weak coupling regime, the cavity mode dispersion remains unchanged, which is revealed by the overlap of PL peak of cavity modes with and without the presence of the SPE. The crossing between SPE and cavity bands is clearly resolved, which is the typical behaviour of weak coupling. In contrast, in a strong coupling regime, the cavity mode (grey dash-line) lies in between UPB (dark-blue dash-line) and LPB (red dash-line), indicating the change of band dispersion from the original cavity mode. Furthermore, the avoided crossing is observed when approaching normal incidence. These are the typical results from emitter-cavity strong coupling.

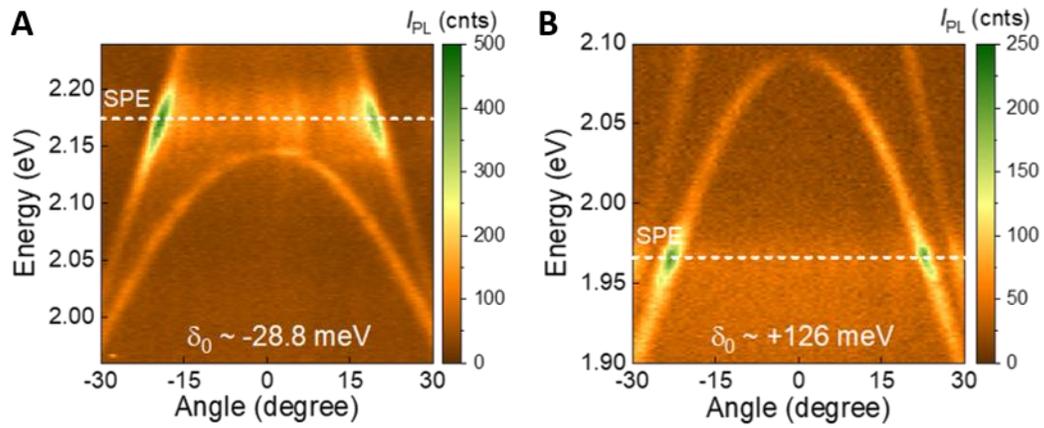

**Fig. S4.** Angle-resolved PL spectra measured for different SPEs on different arrays for (**A**) negative detuning $\delta_0 \sim$ -28.8 meV and (**B**) positive detuning $\delta_0 \sim$ +126 meV. Due to the large energy mismatch, the SPEs only show weak coupling to resonance modes.

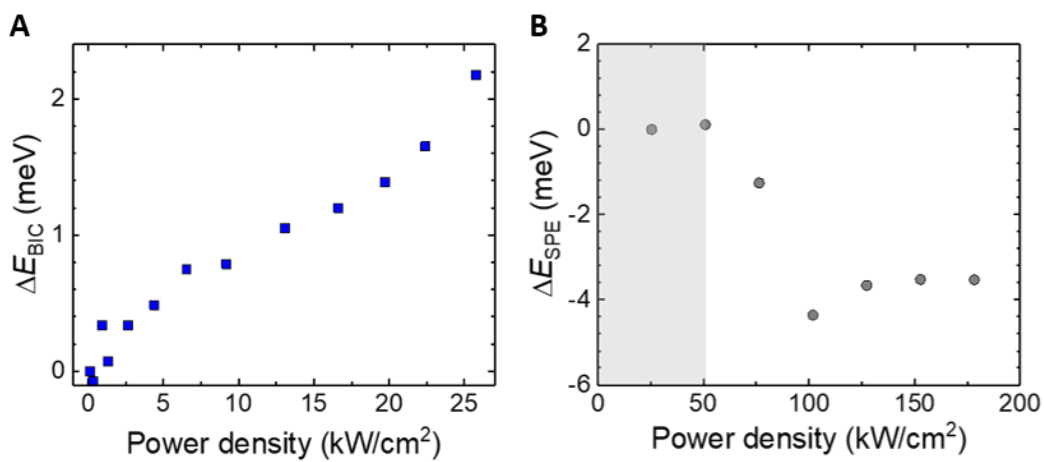

**Fig. S5.** Energy shifts of (**A**) BIC ($\Delta E_{BIC}$) and (**B**) SPE ($\Delta E_{SPE}$) with varying excitation powers. When increasing power density from 0.672 kW/cm² to 25.784 kW/cm², the BIC resonance energy increases by ~2.2 meV, while the SPE peak remains unchanged (grey shaded region).